\documentclass[prr,10pt,superscriptaddress,amsmath,amssymb,aps,physrev]{revtex4-1}
\usepackage{amsmath}
\usepackage{amssymb}
\usepackage{graphicx}
\usepackage{booktabs} 
\usepackage{dcolumn}
\usepackage{braket}
\usepackage{comment}
\usepackage{float}
\usepackage[T1]{fontenc}
\usepackage{mathtools}
\usepackage[titletoc, toc,page]{appendix}

\usepackage[dvipsnames]{xcolor}
\definecolor{dark-blue}{rgb}{0,0.2,0.6}
\usepackage[colorlinks,
            pdfusetitle,
            urlcolor=dark-blue,
            citecolor=dark-blue,
            linkcolor=dark-blue]{hyperref}

\usepackage{etoolbox}
\makeatletter
\pretocmd{\NAT@open}{\begingroup\color{\@citecolor}}{}{}
\apptocmd{\NAT@close}{\endgroup}{}{}
\makeatother

\AtBeginDocument{%
  \heavyrulewidth=.08em
  \lightrulewidth=.05em
  \cmidrulewidth=.03em
  \belowrulesep=.65ex
  \belowbottomsep=0pt
  \aboverulesep=.4ex
  \abovetopsep=0pt
  \cmidrulesep=\doublerulesep
  \cmidrulekern=.5em
  \defaultaddspace=.5em
}

\newcommand{\fixme}[1]{{\color{black} #1}}

\begin{document}

\begin{flushleft}

{\LARGE \textbf{Single-atom imaging of ${}^{173}$Yb in optical tweezers \\[5pt] loaded by a five-beam magneto-optical trap}}

\vspace{1em}

O.~Abdel~Karim\textsuperscript{1,2,*}, 
A.~Muzi~Falconi\textsuperscript{3,*}, 
R.~Panza\textsuperscript{3,1}, 
W.~Liu\textsuperscript{1,\dag}, 
F.~Scazza\textsuperscript{3,1,\ddag}

\vspace{1em}

\textsuperscript{1} Istituto Nazionale di Ottica del Consiglio Nazionale delle Ricerche (CNR-INO), 34149 Trieste, Italy\\
\textsuperscript{2} Department of Physics, University of Naples Federico II, 80138 Naples, Italy\\
\textsuperscript{3} Department of Physics, University of Trieste, 34127 Trieste, Italy\\
\textsuperscript{\dag} Present address: Institute of Laser Spectroscopy, Shanxi University, Taiyuan 030006, China\\
\textsuperscript{*} These authors contributed equally to this work\\
\textsuperscript{\ddag} \texttt{francesco.scazza@units.it}

\vspace{2em}

\end{flushleft}

\noindent
\textbf{Abstract}

\noindent We report on the trapping and imaging of individual ytterbium atoms in arrays of optical tweezers, loaded from a magneto-optical trap (MOT) formed by only five beams in an orthogonal configuration. 
In our five-beam MOT, operating on the narrow ${}^1$S${}_0 \rightarrow {}^3$P${}_1$ intercombination transition, gravity balances the radiation pressure of a single upward-directed beam. This approach enables efficient trapping and cooling of the most common ytterbium isotopes (${}^{171}$Yb, ${}^{173}$Yb and ${}^{174}$Yb) to $\lesssim 20\,\mu$K at densities $\sim 10^{11}$ atoms/cm$^3$ within less than one second. 
This configuration allows for significantly reducing the complexity of the optical setup, potentially benefiting any ytterbium-atom based quantum science platform leveraging single-atom microscopy, from quantum processors to novel optical clocks.
We then demonstrate the first single-atom-resolved imaging of the fermionic, large-spin isotope ${}^{173}$Yb ($I=5/2$), employing a two-color imaging scheme that does not rely on magic-wavelength trapping. We achieve a high single-atom imaging fidelity of $99.96(1)\%$ and a large survival probability of $98.5(2)\%$, despite large differential light shifts affecting all nuclear spin sublevels of the excited ${}^3$P${}_1$ state involved in the cooling transition. The demonstrated capabilities will play a key role in future quantum simulations and computing applications with ${}^{173}$Yb arrays.


\section{Introduction}
Optically trapped alkaline-earth-like atoms (AEAs) have become a leading platform for quantum information processing, quantum simulation, and quantum metrology. Owing to their metastable clock states and the decoupled nuclear spins of fermionic isotopes, AEAs provide robust and versatile qubit encodings \cite{Gorshkov_2009,Shibata_2009,Daley_2011,Pagano_2019,Madjarov_2020,Wu_2022,Chen_2022,Sahay_2023,Jia_2024}, as well as stable frequency references, underpinning state-of-the-art optical clocks \cite{Ludlow_2015,McGrew_2018,Oelker_2019,Zheng_2022} and matter-wave interferometers \cite{Hu_2017,Rudolph_2020}. These features also make ultracold AEAs an unparalleled resource for the simulation of novel quantum many-body systems \cite{Gorshkov_2010,Cazalilla_2014, Zhang_2020,Surace_2023,Surace_2024}. In particular, the fermionic isotope ${}^{173}$Yb --- with its nuclear spin $I=5/2$ and strong magnetic interactions between clock states --- gives access to highly symmetric SU($N$) models of orbital magnetism \cite{Gorshkov_2010,Cazalilla_2014,Scazza_2014,Zhang_2014,Cappellini_2014,Riegger_2018,KanaszNagy_2018,Zhang_2020,Abeln_2021}, largely unexplored to date.

To fully exploit the potential of AEAs for quantum science applications, it is crucial to develop scalable and reliable methods for efficient atom cooling and trapping, compatible with high-resolution detection at the single-particle level. In recent years, optical tweezer arrays have enabled significant advances in the microscopic control and detection of individual AEAs \cite{Cooper_2018,Norcia_2018,Saskin_2019,Covey_2019,Jenkins_2022,Ma_2022,Huie_2023,Norcia_2023,Nakamura_2024,Tao_2024}, facilitating striking demonstrations of high-fidelity quantum operations \cite{Madjarov_2020,Ma_2023,Scholl_2023,Lis_2023} and novel tweezer clock architectures \cite{Madjarov_2019,Norcia_2019,Young_2020,Cao_2024,Shaw_2024,Finkelstein_2024}. Two isotopes of ytterbium, bosonic ${}^{174}$Yb and fermionic ${}^{171}$Yb, have been previously trapped in optical tweezer arrays and imaged with high single-atom detection fidelity \cite{Saskin_2019,Jenkins_2022,Ma_2022} --- a key capability for quantum processors and simulators \cite{Gross_2017,Browaeys_2020,Kaufman_2021}. On the other hand, fermionic ${}^{173}$Yb atoms have yet to be individually trapped or imaged. 

In experiments with AEAs, atoms are loaded into optical tweezers from a cold atomic sample in a narrow-line magneto-optical trap (MOT), which conventionally utilizes a six-beam configuration requiring optical access along all three axes. This setup often conflicts with the high numerical aperture (NA) required from at least one side of the atomic sample, where one or more microscope lenses are positioned to allow for addressing and imaging atoms with sub-micron optical resolution. In this work, we demonstrate the loading of ytterbium atoms into optical tweezers from a MOT operating with five beams in an open-top configuration, where the downward beam is omitted and entirely replaced by gravity to provide the necessary confinement along the vertical direction. This approach, demonstrated so far for dysprosium and erbium \cite{Ilzhofer_2018}, eliminates the need for more intricate beam arrangements, simplifying the experimental setup while maintaining high loading efficiency \cite{Grun_2024}. It is particularly advantageous for compact, tweezer-based quantum science platforms with single-atom resolution, where a microscope objective is essential.

We show that such five-beam configuration allows to achieve the density and temperature necessary to efficiently load an optical tweezers array with any of the most common ytterbium isotopes --- $^{174}\text{Yb}$, $^{171}\text{Yb}$, and $^{173}\text{Yb}$. 
Further, we demonstrate the high-fidelity imaging of individually trapped fermionic $^{173}$Yb atoms, exploiting a two-color scheme to simultaneously yield sufficient cooling and photon scattering rates \cite{Yamamoto_2016, Cooper_2018, Jenkins_2022, Urech_2022}, even in the presence of large tweezer-induced differential light shifts at our trapping wavelength of $532$\,nm. 
Despite the complex internal structure of $^{173}$Yb and our peculiar five-beam molasses geometry, we achieve a high detection fidelity $>99.95\%$ and low losses $\sim1.5\%$, demonstrating single-atom imaging performances comparable to those obtained for $^{171}$Yb in similar trapping conditions using a conventional six-beam configuration \cite{Jenkins_2022}. 
Our technique could be extended to other high-spin AEAs, such as $^{87}$Sr, possibly employing Sisyphus cooling instead of Doppler molasses to address a subset of the spin states. 

\begin{figure}[t]
    \centering    \includegraphics[scale=0.5]{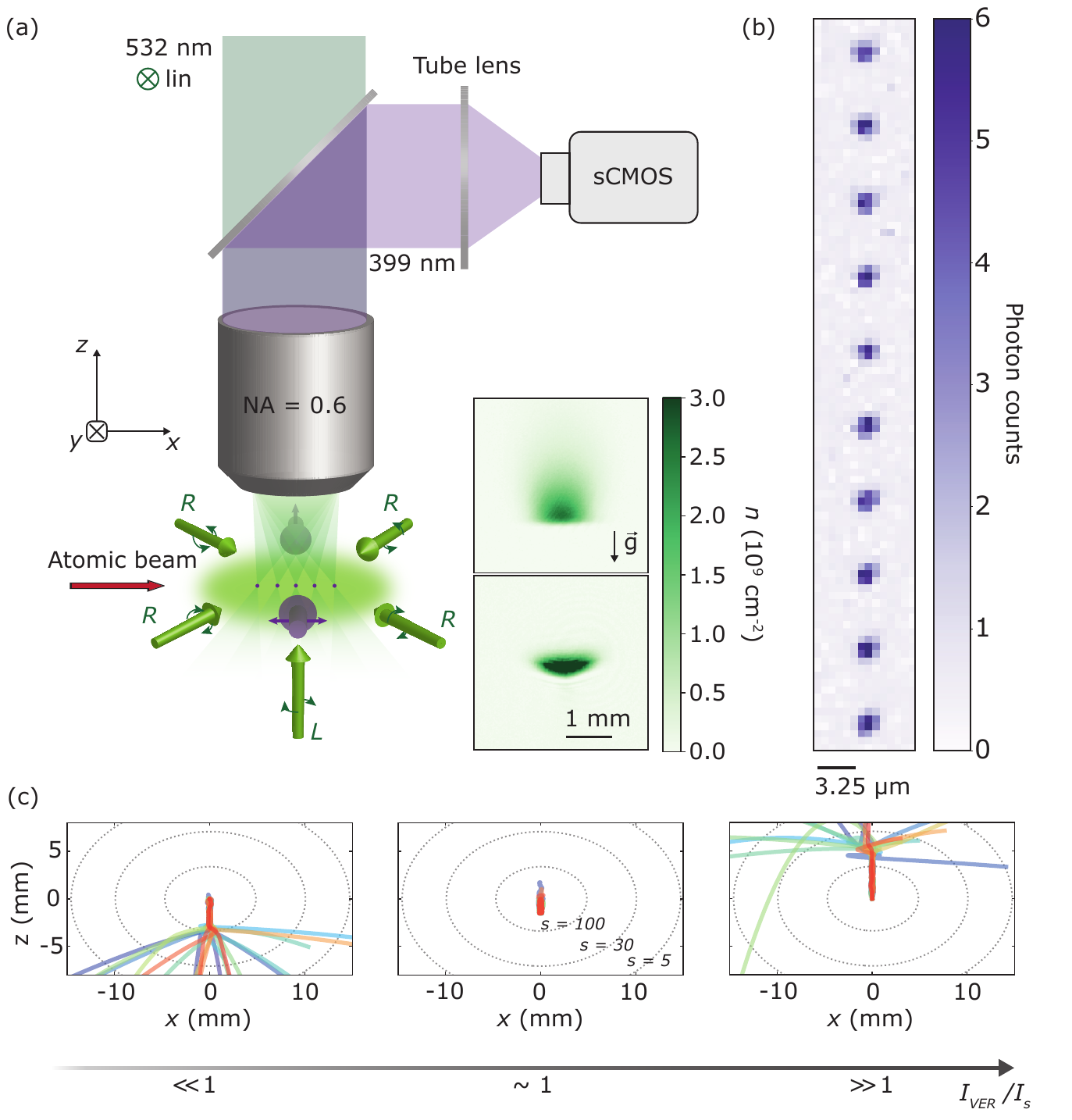}
    \caption{Overview of the five-beam MOT and trapping of ytterbium atoms in optical tweezers. (a) Narrow-line MOT operating on the ${}^1$S$_0 \rightarrow {}^3$P$_1$ transition at $556\,\text{nm}$ in a five-beam configuration, where the top beam along gravity is omitted. The insets display absorption images of a $^{174}\text{Yb}$ MOT (top, $2\,\text{s}$ loading) and cMOT (bottom, $0.7\,\text{s}$ loading). Atoms are loaded into the optical tweezers array directly from the cMOT. A $0.6$ NA objective is used to focus tweezer traps at $532\,\text{nm}$ and to collect the atomic fluorescence on the ${}^1$S$_0 \rightarrow {}^1$P$_1$ transition at $399\,\text{nm}$. A single horizontal retroreflected beam at $399\,\text{nm}$, \fixme{propagating along the $y$-axis}, is employed to excite atoms, while the MOT beams are used to cool atoms during the imaging. (b) Average of $300$ images of a tweezer-trapped $^{174}\text{Yb}$ atom array. (c) Monte Carlo simulations of single-atom trajectories in the five-beam MOT. Atoms are initialized at the intersection of the MOT beams and the dynamics is tracked for $200$\,ms. The MOT beam waists are $8\,\text{mm}$ and $5\,\text{mm}$ in the horizontal and vertical directions, respectively. Tuning the vertical beam intensity $I_\mathrm{VER}$, we identify a regime where the average radiation pressure balances gravity holding the atoms without pushing them out of the MOT region. Dotted ellipses represent the intensity profile of one of the horizontal beams, described by the saturation parameter $s=I_\mathrm{HOR}/I_s$.}
    \label{figure_1}
\end{figure}

An overview of our experimental setup is provided in Fig.~\ref{figure_1}(a). Within an octagonal glass cell, a five-beam (5B) MOT operating on the narrow-line ${}^1$S$_0 \rightarrow {}^3$P$_1$ transition at $556\,\text{nm}$ ($\Gamma_\mathrm{556}/2\pi=182\,\text{kHz}$) is loaded in $200\,\text{ms}$. Our MOT is loaded by an ytterbium atomic beam with average longitudinal velocity of about $40$\,m/s (see Appendix~\ref{App:A} for more details on the atomic source). We employ two crossed beams (CBs) acting on the blue ${}^1$S$_0 \rightarrow {}^1$P$_1$ transition at $399\,\text{nm}$ ($\Gamma_\mathrm{399}/2\pi=29\,\text{MHz}$) to further decelerate atoms below the narrow-line MOT capture velocity \cite{Plotkin-Swing_2020, Lunden_2020, Seo_2020}, avoiding the need of a broad-line MOT which is incompatible with a 5B configuration (see Section~\ref{Sec:5B}). 
\fixme{Indeed, a broad-line transition lacks the velocity selectivity needed to hold atoms against gravity without pushing those already moving upward out of the trap.}
The CBs cross \fixme{on the horizontal $xy$-plane, approximately} $22\,\text{mm}$ upstream from the MOT center, and are angled by $21^\circ$ relative to the atomic beam direction.
After loading, the MOT is compressed in $250\,\text{ms}$ to increase its density and reduce its temperature before loading the optical tweezer array \fixme{(see Appendix~\ref{App:A})}. 
The insets in Fig.~\ref{figure_1}(a) show the typical column densities of the 5B MOT and compressed MOT (cMOT) of $^{174}$Yb, obtained via absorption imaging. 

The one-dimensional tweezer array is generated using an acousto-optic deflector (AOD) working at $532\,\text{nm}$. We are able to generate up to $40$ optical tweezers with an inhomogeneity below $2\%$ at a minimum spacing around $ 1.3\,\mu\text{m}$ (see Appendix~\ref{App:C}).
In typical experiments, we load tweezer arrays with a trap depth of about $2$\,mK and a spacing of $8.7\,\mu\text{m}$ from the 5B cMOT. The optical tweezer light is linearly polarized along the $y$-axis, and is focused onto the cold cMOT cloud by a custom microscope objective with $\text{NA}\simeq 0.6$ and an effective focal length of approximately $26\,\text{mm}$.
Single-atom imaging is performed by illuminating the atoms with retroreflected, linearly polarized $399$\,nm light, \fixme{with the forward beam polarized along the $x$-axis and the backward beam along the $z$-axis.} 
\fixme{During imaging, a small magnetic field is applied along the $y$-axis ($B_y = 0.2\,\text{G}$).}
The atomic fluorescence is collected using the same objective employed to focus the tweezer traps. The tweezer light and the atomic fluorescence are decoupled at the back of the objective using a dichroic mirror that transmits light at $532\,\text{nm}$ and reflects light at $399\,\text{nm}$. The atomic fluorescence is then imaged onto a \mbox{sCMOS} camera (Teledyne Photometrics Kinetix22) using a $200\,\text{mm}$ tube lens. The 5B MOT beams are turned on during imaging to provide molasses cooling of atoms trapped in the tweezer array (see Section~\ref{Sec:imaging}).
Figure~\ref{figure_1}(b) shows a typical average image of a $10$-tweezer array of $^{174}$Yb, obtained from approximately $300$ single-shot images. 

\section{Five-beam MOT}\label{Sec:5B}
Compared to a conventional six-beam MOT, in a 5B configuration the down-propagating vertical beam is omitted and the upward push from the bottom beam is counteracted by gravity alone. This scheme leverages the interplay between narrow-line photon scattering and gravity, making it particularly effective for heavy atoms, such as lanthanides \cite{Ilzhofer_2018}.
To gain insight into the 5B MOT working principle and verify its feasibility for ytterbium, we perform Monte Carlo simulations of the atomic trajectories in the MOT (see also Appendix~\ref{App:B}). Trajectories are divided in time steps. For each step, we determine the photon scattering probability, taking into account the magnetic-field gradient, the detuning from resonance and the intensity profile of each MOT beam. Atomic coordinates are then updated considering the combined effects of gravity and stochastic single photon scattering events. Figure~\ref{figure_1}(c) displays $20$ single-atom trajectories in three scenarios differing from one another by the \fixme{intensity of the vertical beam propagating along the $z$-axis}. Atoms start at the intersection of the MOT beams, i.e.~at the zero of the quadrupole magnetic field, and the trajectory is tracked for $200$\,ms. For $I_\mathrm{VER} \ll I_s$, where $I_s \simeq 0.139$\,mW/cm$^2$ is the saturation intensity of the ${}^1$S$_0 \rightarrow {}^3$P$_1$ transition, the \fixme{vertical} beam intensity is insufficient to hold atoms against gravity, causing them to fall downwards out of the MOT region. On the other hand, when $I_\mathrm{VER} \gg I_s$ the atoms escape upwards due to the excessive radiation pressure from the bottom beam. 
At $I_\mathrm{VER} \sim I_s$, the mean force from the upward beam balances gravity, allowing atoms to remain trapped in the MOT. This condition is achieved at an intensity around $I_s$ as an outcome of the small linewidth of the ${}^1$S$_0 \rightarrow {}^3$P$_1$ transition and the large mass of ytterbium. Indeed, lighter atoms would experience a stronger acceleration, thus requiring a weaker intensity, resulting in reduced robustness. 
\begin{figure}[t]
    \centering
    \includegraphics[scale=0.5]{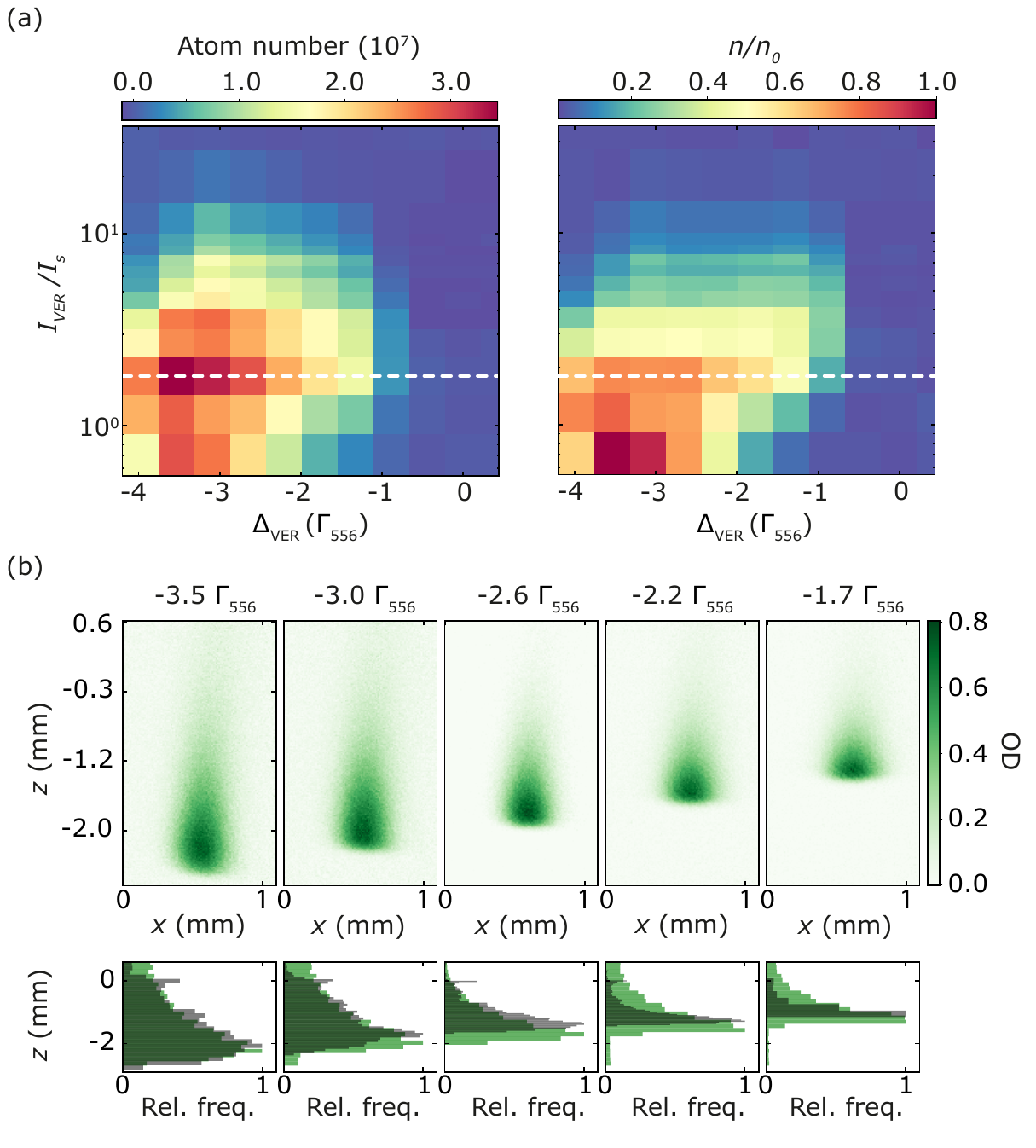}
    \caption{Dependence of $^{171}$Yb five-beam MOT loading on the vertical beam parameters. (a) Atom number captured by a $^{171}$Yb 5B MOT (left) and normalized atom peak density $n/n_0$ (right) after $2\,\text{s}$ loading time as a function of the \fixme{intensity of the vertical beam} $I_\mathrm{VER}$ and detuning $\Delta_\mathrm{VER}$. The magnetic field gradient is set to $5\,\text{G/cm}$, while the horizontal beams have a detuning of $\Delta \simeq -3.4\,\Gamma_{556}$ and an intensity of $I_\mathrm{HOR} \simeq 60\,I_s$ \fixme{for each beam}. (b) Top: in situ absorption images of the 5B MOT for different detunings of the vertical beam at fixed intensity $I_\mathrm{VER} \simeq 2 I_s$, corresponding to the dashed line in panel (a). The cloud shifts downwards as the detuning moves away from resonance. Bottom: 
    histograms of the experimental optical density (green) and of the simulated trajectories (grey) sampled on the $z$-axis. Each simulated histogram is the result of $20$ independent trajectories. The zero of the $z$-axis in experimental images is set to optimally align them with simulations.}
    \label{figure_2}
\end{figure}
 
In a 5B MOT, unlike in a conventional six-beam (6B) MOT, the primary role of the \fixme{vertical} beam is to counteract gravity, while the \fixme{horizontal} beams \fixme{($xy$-plane)} must slow and capture the horizontally incoming atoms. Consequently, a 5B MOT greatly benefits from an unconstrained control on the \fixme{horizontal} and \fixme{vertical} beam intensities and detuning.
Figure~\ref{figure_2}(a) illustrates the effect of the intensity $I_\mathrm{VER}$ and vertical beam detuning $\Delta_\mathrm{VER}$ on the trapped number of atoms and the peak atomic density measured in a 5B MOT of $^{171}$Yb after $2$\,s of loading. 
Consistently with the simulated atomic trajectories in Fig.~\ref{figure_1}(c), high intensities of the vertical beam lead to a lower number of captured atoms as they escape from the MOT vertically. Similarly, when the detuning is too close to resonance, the resulting scattering force becomes too strong, causing atoms to be pushed out of the MOT region. Notably, the atom number and peak density do not reach their maxima for the same conditions, since variations in the vertical beam parameters also affect the shape of the atomic cloud. In the top panel of Fig.~\ref{figure_2}(b), we present absorption images of the cloud for a fixed vertical beam intensity $I_\mathrm{VER} \simeq 2\,I_s$, corresponding to the dashed horizontal line in Fig.~\ref{figure_2}(a). For decreasing detuning from resonance, the cloud is pushed upwards in an unstable region above the center of the quadrupole field, leading to a reduced atom number. 
By increasing the detuning, on the other hand, the cloud shifts downwards as atoms redistribute to regions of the magnetic field gradient where the radiation pressure from the upward beam balances gravity. At the same time, the cloud develops into a drop-like shape, maximizing the total atom number while maintaining a nearly constant peak density. This behavior is confirmed in Monte Carlo simulations. In the bottom panel of Fig.~\ref{figure_2}(b) we compare the measured horizontally-integrated atomic densities with simulated distributions, obtained by binning the $z$-positions of $20$ independent atomic trajectories for each detuning. We can approximate that each atom samples all accessible states, since the atomic motion \fixme{proceeds on timescales} on the order of $h/E_\mathrm{rec} \simeq 270\,\mu$s (where $E_\mathrm{rec}$ is the atomic recoil energy from a $556$\,nm photon), much shorter than the total simulation time of $200$\,ms. We thus compare the entire simulated trajectory of a single atom with an experimental image of the MOT cloud observed at a specific time. 
The simulations reproduce both the vertical shift of the atomic cloud position and the asymmetric density profile broadening as a function of the vertical beam detuning.

\begin{figure}[t]
    \centering
    \includegraphics[scale=0.5]{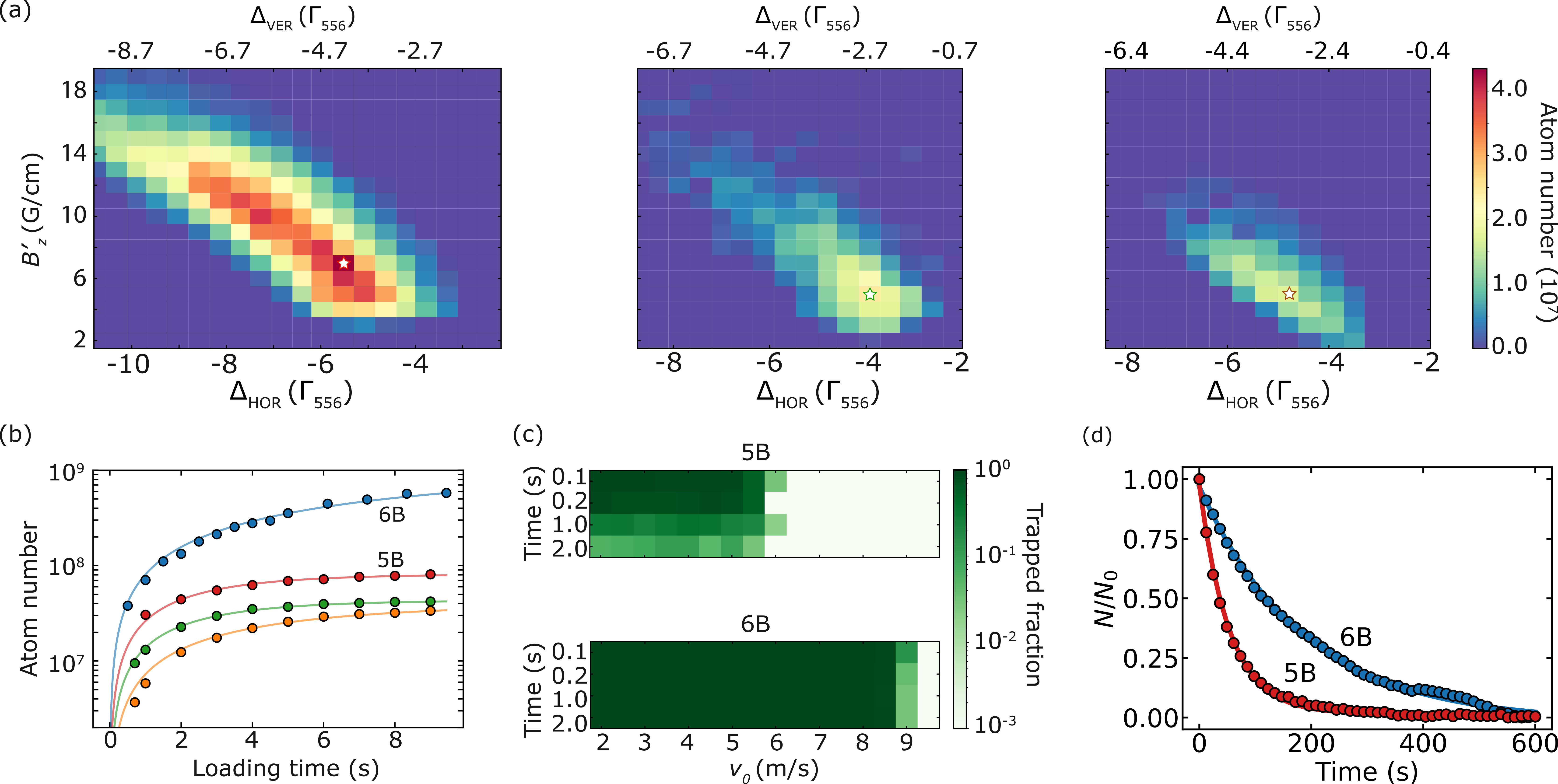}
    \caption{Five-beam MOT loading and stability for bosonic and fermionic isotopes. (a) Atom number $N$ as a function of horizontal and vertical beam detunings and magnetic-field gradient for $^{174}\text{Yb}$ (left), $^{171}\text{Yb}$ (center) \fixme{and $^{173}\text{Yb}$ (right)} after $2\,\text{s}$ of loading. \fixme{The intensities of horizontal beams and the vertical beam} are $220\,I_s$ and $3\,I_s$ for $^{174}\text{Yb}$, $60\,I_s$ and $4\,I_s$ for $^{171}\text{Yb}$ \fixme{and $200\,I_s$ and $6\,I_s$ for $^{173}\text{Yb}$} respectively.
    (b) Loaded atoms as a function of time in a $^{174}\text{Yb}$ 6B MOT (blue circles) and a 5B MOT of $^{174}\text{Yb}$ (red circles), $^{171}\,\text{Yb}$ (green circles) \fixme{and $^{173}\,\text{Yb}$ (orange circles)}. For $^{174}$Yb 6B MOT, the magnetic-field gradient and detuning $\Delta_\mathrm{HOR}$ are set to $7\,\text{G/cm}$ and $-4\,\Gamma_{556}$, respectively. The values used for the 5B MOTs are marked by star symbols in panel (a). Solid lines represent exponential fits to the data (see text). (c) Monte Carlo simulations of the capture and stability of a 5B and 6B MOT. We simulate the capture process for $100$ independent atoms, entering the MOT region with variable velocity $v_0$. The time-dependent captured fraction is determined as the number of atoms remaining in the MOT region after a variable holding time. The 5B and 6B simulations are performed using different beam intensities and detunings, optimizing the capture in each of the two cases. (d) MOT relative atom number $N/N_0$ as a function of holding time for $^{174}\text{Yb}$, obtained in the 6B (blue circles) and 5B (red circles) configurations. Solid lines are exponential fits, 
    yielding a lifetime $\tau_{6B}=185(4)\,\text{s}$ and $\tau_{5B}=54(1)\,\text{s}$ for the two configurations, respectively.}
    \label{figure_3}
\end{figure}

Figure~\ref{figure_3}(a) reports the measured number of trapped atoms as a function of the overall beam detuning and magnetic-field gradient for $^{174}\text{Yb}$, $^{171}\text{Yb}$ \fixme{and $^{173}\text{Yb}$} after $2\,\text{s}$ of loading. 
\fixme{The loading of the fermionic isotopes displays reduced robustness, which we attribute to the hyperfine and nuclear sub-structure of the ${}^1$S$_0 \rightarrow {}^3$P$_1$ transition.} 
\fixme{In particular, the probability of scattering photons on the stretched $|F, m_F=\pm F\rangle\rightarrow |F'=F+1, m_F'=\pm F' \rangle$ transitions is reduced with respect to the case of $^{174}$Yb~($I=0$).} 
\fixme{Moreover, the presence of nearly degenerate $\sigma^\pm$-transitions such as $m_F=\pm1/2 \rightarrow m_F'=\mp1/2$ leads to position-dependent competing optical forces~\cite{Mukaiyama_2003,Stellmer_2014}. These  include anti-restoring forces that push atoms out of the trap. This effect is especially pronounced in $^{173}$Yb, where the large nuclear spin results in a more complex hyperfine structure and a denser manifold of near-degenerate $\Delta m_F = \pm1$ transitions with respect to $^{171}$Yb. The issue is well-known in narrow-line MOTs of ${ }^{87}$Sr~($I=9/2$), where a so-called stirring beam is typically employed to minimize its negative  impact~\cite{Mukaiyama_2003,Stellmer_2014}}. In Fig.~\ref{figure_3}(b), we compare the loading of a $^{174}\text{Yb}$ 6B MOT with that of $^{174}\text{Yb}$, $^{171}\text{Yb}$ \fixme{and $^{173}\text{Yb}$} 5B MOTs. The loaded atom number $N$ as a function of time is fitted with an exponential function, $N(t) = N_s \left(1 - e^{-t/\tau_L}\right)$, where $N_s$ represents the stationary loaded atom number and $\tau_L$ is the mean loading time. We can then define a loading rate, $R = N_s/\tau_L$, which quantifies the efficiency of atom collection. Table~\ref{table_figure_2} reports the values of $N_s$, $\tau_L$, and $R$ for different isotopes and MOT configurations.
We find that the 5B MOT atom number saturates faster and at a lower value $N_s$ than for the 6B MOT, which indicates a reduced stability of the 5B configuration, irrespective of the specific isotope. 
\fixme{Additionally, the $^{171}\text{Yb}$ and $^{173}\text{Yb}$ 5B MOTs exhibit a lower stationary atom number $N_s$ compared to $^{174}\text{Yb}$, which is consistent with their reduced isotopic abundances. The loading rate $R$ of $^{173}\text{Yb}$ is smaller than that of $^{171}\text{Yb}$ by a factor of approximately 2, possibly associated also with inefficiencies in the 2D MOT and CBs slowing stemming from its particularly complex hyperfine structure. In particular, the ${}^1$S$_0 \rightarrow {}^1$P$_1$ transition is characterized by a small hyperfine shift of approximately 70\,MHz between the $F'=3/2$ and $F'=7/2$, which complicates efficient 2D MOT operation and CBs slowing on the red-detuned side of the ${}^1$S$_0\,|F=5/2\rangle \rightarrow {}^1$P$_1\,|F'=7/2\rangle$ transitions.}
For all isotopes, we ascribe the reduced performance of the 5B MOT to an intrinsic residual loss mechanism, associated with the presence of an escape channel for atoms along the vertical direction. Specifically, fluctuations in the upward scattering force lead to atoms exiting the trapping region vertically and being lost \footnote{Fluctuations of the radiation force from the vertical upward beam are enhanced by the optical pumping processes within the large nuclear spin structure of $^{173}\text{Yb}$, where the transitions associated with each nuclear spin state of the ground manifold feature widely different line strengths.}. Consequently, the 5B MOT reaches saturation in a shorter time $\tau_L$ with respect to the 6B configuration. 

To investigate this behavior in more detail, we simulate the trajectories of $100$ independent atoms entering the MOT region with different longitudinal velocities $v_0$, and we determine the number of atoms remaining trapped in the MOT after a variable time from the start of the simulation. In simulations, we fix the beam parameters and magnetic-field gradient to optimal values, similar to the ones found in the experiment. Results for holding times comparable with the typical experimental loading times are shown in Fig.~\ref{figure_3}(c). For the same horizontal beam waist of $8$\,mm, the 5B MOT exhibits a lower capture velocity of about $5\,\text{m/s}$ compared to $8.5\,\text{m/s}$ of the 6B configuration. Moreover, unlike in the 6B MOT, the trapped fraction in the 5B MOT decays over time. 
In the 5B case, the reduced vertical beam intensity leads to preferential scattering from the horizontal beams (see Appendix~\ref{App:B}). When the horizontal intensity is too high, this imbalance causes atoms to fall out of the trap. To improve stability, the horizontal beam intensity must thus be reduced compared to the 6B case, resulting in a lower capture velocity. 
The reduced stability of the 5B MOT is thus confirmed in the individual trajectory approach of our simulations, which do not include the continuous loading of atoms during the holding time. The MOT stability can also be investigated by measuring its lifetime once the inflow of atoms from the atomic beam has been extinguished. 
In Fig.~\ref{figure_3}(d), we compare the lifetime measured for $^{174}$Yb 5B and 6B MOTs. The fluorescence signal from the MOT is recorded over time using a CMOS camera, and the data is fitted with an exponential decay, yielding lifetimes $\tau_{6B} = 185(4)\,\text{s}$ and $\tau_{5B} = 54(1)\,\text{s}$. While the reduction of the lifetime in the 5B configuration is significant, the lifetime remains sufficiently large --- so as not to pose any issue for the following steps in the experimental sequence. 

\begin{table}[t!]
\bigskip
	\centering
	\small 
	\setlength{\tabcolsep}{8pt}
	\begin{tabular}{lc|c|c|c}
		\toprule
		& \multicolumn{1}{c|}{6B ($^{174}$Yb)} & \multicolumn{1}{c|}{5B ($^{174}$Yb)} & \multicolumn{1}{c|}{5B ($^{171}$Yb)} & \multicolumn{1}{c}{5B ($^{173}$Yb)} \\ 
		\midrule
		\midrule
		$\tau_L\,(\mathrm{s})$        			 & 14(1)    & 2.6(2) & 2.60(7) & 5.0(3) \\
		$N_{s}\,(10^8)$		 & 11.9(7)  & 0.81(2) & 0.433(3) & 0.40(1) \\
            $R\,(10^8 \,\mathrm{s}^{-1})$    & 0.84(5)	& 0.32(1) & 0.166(1) & 0.080(2)\\
		\bottomrule
	\end{tabular}
	\caption{MOT loading rate parameters. Mean loading time $\tau_L$, stationary atom number $N_{s}$ and loading rate $R$, obtained from exponential fits of $N(t)$ for
    different isotopes and beam configurations. \fixme{The temperature of the atomic source oven was kept constant across  measurements with different isotopes.}}
	\label{table_figure_2}
\end{table}

After loading, the 5B MOT is compressed in $250\,\text{ms}$ to enhance the atomic density and reduce the temperature to values suitable for efficiently loading optical tweezer traps. \fixme{For this, the MOT beams intensities and detuning from resonance are reduced, while the magnetic quadrupole field is increased (see Appendix~\ref{App:A}).} During the compression, the center of the quadrupole field is gradually adapted using a vertical pair of bias coils to keep the 5B MOT fixed to its original position, despite the change of vertical beam intensity and detuning. We realize a 5B cMOT for $^{174}$Yb, $^{171}$Yb, and $^{173}$Yb, reaching temperatures of $21(6)\,\mu\text{K}$, $23(5)\,\mu\text{K}$, and $9(2)\,\mu\text{K}$, respectively, with typical mean densities around $10^{11}\,\text{atoms cm}^{-3}$ for all isotopes. The obtained temperatures for all isotopes are close to the Doppler temperature limit, $T_D = \hbar \Gamma_\mathrm{556}/(2k_B)\simeq 4.4\,\mu$K. The lower measured temperature for $^{173}$Yb is likely the outcome of significant sub-Doppler cooling effects, associated to its high angular momentum structure~\cite{Maruyama_2003}.

\section{Loading and imaging single atoms in tweezer arrays}\label{Sec:imaging}
\begin{figure}[t!]
    \centering
    \includegraphics[scale=0.5]{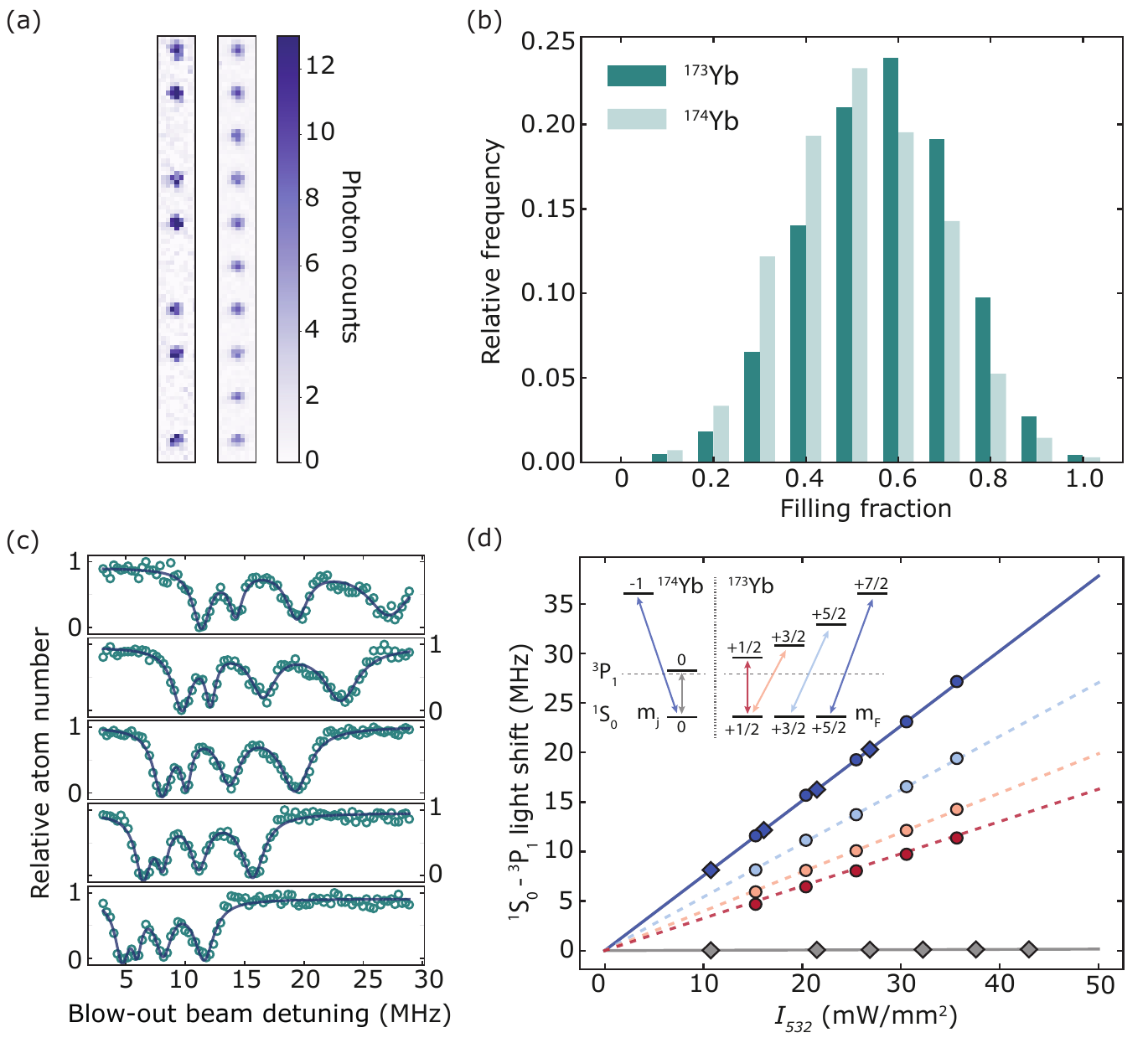}
    \caption{Loading and imaging of $^{173}$Yb arrays. (a) Single-shot and average of more than $2500$ images of a $10$-tweezer array of $^{173}$Yb. The imaging pulse duration and intensity are set to $50\,\text{ms}$ and $I_{399}\sim 10^{-2}\,I_s $, respectively. (b) Filling fraction of the tweezer array, obtained from over $2500$ images for both $^{174}$Yb and $^{173}$Yb. A slight enhancement in the filling fraction for $^{173}$Yb is noticeable, with an average of $58\%$ compared to $50\%$ for $^{174}$Yb. (c) Blow-out spectroscopy of the  ${}^1$S$_0\, |F=5/2\rangle \rightarrow {}^3$P$_1\, |F^\prime=7/2\rangle$ transition for tweezer-trapped $^{173}$Yb atoms. The normalized atom number is plotted as a function of the blow-out pulse detuning from the free-space resonance, for varying tweezer trap depths [top to bottom: $1.8,\,1.5,\,1.3,\,1.0,\,0.8\,\text{mK}$]. The spectroscopic signal is fitted with a sum of four Lorentzian functions (solid lines) to determine the $m_F'$-dependent differential light shift. (d) Differential ${}^1$S$_0 - {}^3$P$_1$ light shift for $^{174}$Yb (diamonds) and $^{173}$Yb (circles) as a function of the $532$\,nm tweezer peak intensity $I_\mathrm{532}$. Solid lines are linear fits of the $^{174}$Yb experimental data. 
    Dashed lines are the predicted $^{173}$Yb light shifts obtained from the measured $^{174}$Yb light shifts (see text). 
    The inset displays a scheme of the transitions associated to the measured light shifts with corresponding colors. Although all allowed transitions from each ground-state $m_F$ sub-level are coupled by the blow-out pulse, only a few are depicted to avoid clutter. Grey dashed lines represent the free space energy of the excited states.}
    \label{figure_4}
\end{figure}
\begin{figure*}[t!]
    \centering
    \includegraphics[width=0.85\textwidth]{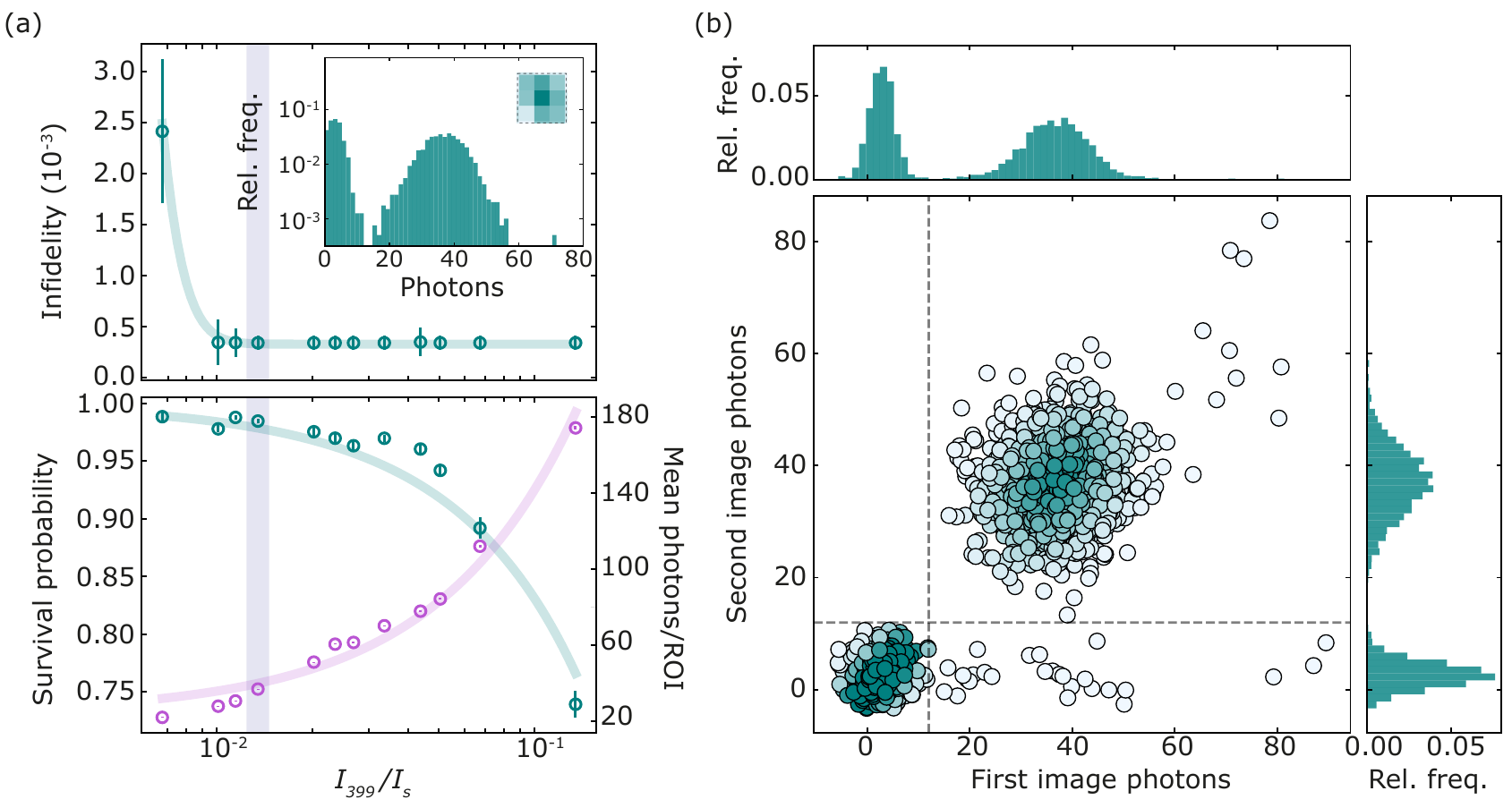}
    \caption{Single-atom imaging of $^{173}$Yb. (a) Imaging infidelity, survival probability and mean photon number per ROI as function of imaging beam intensity.  Infidelity and survival probability data points are obtained from two consecutive imaging pulses of $50\,\text{ms}$ duration, separated by $50\,\text{ms}$ dark time. The vertical shaded line indicates the intensity at which $99.96(1)\%$ fidelity, $98.5(2)\%$ survival probability, and $36.8(1)$ mean photons/ROI are obtained. Shaded lines under the data are guides to the eye. The inset shows the photon count histogram for the first image, integrating over the 3x3 pixel ROIs centered at each trap.
    \fixme{Each point is obtained from about $300$ images of a $10$-tweezer array.}
    (b) Correlation between photon counts in the first and second images at the intensity marked by the shaded line in (a). 
    The gray dashed lines indicate the photon count threshold which maximizes the global fidelity.}
    \label{figure_5}
\end{figure*}

We demonstrate the loading and high-fidelity imaging of single atoms in optical tweezers for bosonic and fermionic ytterbium isotopes. Here, we focus on $^{173}$Yb, for which single-atom imaging has not been previously demonstrated.
We load a $10$-tweezer array at $532\,\text{nm}$ from the 5B cMOT by turning on the traps during the whole MOT compression stage. Optical tweezers have a beam waist ($1/e^2$-radius) of approximately $580$\,nm (see Appendix~\ref{App:D}) and a trap depth $\simeq 2\,\text{mK}$. 
To achieve stochastic single-atom occupations, we induce light-assisted collisions (LACs) and consequent pairwise atom ejection \cite{Jenkins_2022, Grunzweig}. In particular, LACs are performed by applying a $50\,\text{ms}$ red-detuned $556$\,nm pulse after loading the traps. 
To image single atoms, we employ a two-color scheme \cite{Yamamoto_2016, Cooper_2018, Jenkins_2022, Urech_2022}. We drive weakly the ${}^1$S$_0 \rightarrow {}^1$P$_1$ transition for imaging, while simultaneously cooling on the ${}^1$S$_0 \rightarrow {}^3$P$_1$ transition to preserve the atom inside the trap \cite{Yamamoto_2016, Saskin_2019, Jenkins_2022}.
Imaging typically takes tens of milliseconds, determined by the interplay between the two mechanisms. As the linewidths of the two transitions are widely different ($\Gamma_{399}/\Gamma_{556} \sim 167$), the intensity of the imaging light must be sufficiently low to limit the photon scattering rate and the resulting heating. 
Figure~\ref{figure_4}(a) presents a single-shot image and an average over multiple realizations of a $^{173}\text{Yb}$ tweezer array. Images are obtained with an imaging pulse duration of $50\,\text{ms}$ and an imaging beam intensity $I_{399} \simeq 10^{-2}\,I_s$,  where $I_s \simeq 60$\,mW/cm$^2$ is the saturation intensity of the ${}^1$S$_0 \rightarrow {}^1$P$_1$ transition.
Histograms in Fig.~\ref{figure_4}(b) show the single-atom loading probability distributions for $^{174}$Yb and $^{173}$Yb. We observe a slightly higher loading efficiency for $^{173}$Yb, achieving a mean filling fraction $\simeq 58\%$, to be compared with $\simeq 50\%$ for $^{174}$Yb. This suggests that for $^{173}$Yb an enhanced light-assisted loading mechanism is present~\cite{Grunzweig,Brown_2019,Jenkins_2022,Huie_2023}.

As our imaging scheme relies on narrow-line cooling, it would be advantageous to work under magic, i.e.~state-independent, trapping conditions for the two states involved in the cooling transition \cite{Saskin_2019, Yamamoto_2016, Covey_2019}. In non-magic traps, a non-zero differential light shift is instead present between the two states, potentially compromising the effectiveness of narrow-line cooling: the frequency of the cooling transition depends the location of the atom within the trap and thus on the atom's energy. 
In addition, the transition frequency depends on the tweezer depth, and tweezer \fixme{intensity} inhomogeneities across the array lead to reduced cooling performances. 
Linearly-polarized $532$\,nm traps were previously found to be magic for the ${}^1$S$_0 \ket{\, J=0, m_J=0} \rightarrow {}^3$P$_1 \,\ket{J'=1, m_J^\prime=0}$ transition of $^{174}$Yb \cite{Saskin_2019}. However, none of the corresponding ${}^1$S$_0 \ket{\, F=5/2, m_F} \rightarrow {}^3$P$_1 \ket{\,F^\prime=F\pm 1, m_F^\prime}$ transitions in $^{173}$Yb are predicted to feature a vanishing differential light shift at such wavelength. 

To minimize array inhomogeneities, we apply a feedback protocol to our tweezer array~\cite{Jenkins_2022}, achieving an intensity inhomogeneity of less than $1.2\%$ across a $10$-trap array (see Appendix~\ref{App:C} for details). To characterize the trapping condition for $^{173}$Yb, we measure the differential light-shift between the $^1$S$_0$ and $^3$P$_1$ states. To this aim, for both $^{174}$Yb and $^{173}$Yb, we perform spectroscopy with a green blow-out pulse (containing all polarizations) that expels atoms from the traps prior to imaging. 
Typical spectra of $^{173}$Yb atoms are shown for different trapping intensities in Fig.~\ref{figure_4}(c), where the measured relative atom number is plotted as a function of the blow-out pulse detuning from the free-space ${}^1$S$_0\,|F=5/2\rangle \rightarrow {}^3$P$_1\,|F'=7/2\rangle$ resonance. Here, the tweezer polarization is aligned along the quantization B-field, suppressing any vector light shift but not the tensor light shift, characterized by a $|m_F'|^2$ dependence \cite{Steck_2011}. We indeed observe four spectroscopic peaks corresponding to the different Zeeman sub-levels of the $^3$P$_1$~$|F'=7/2\rangle$ hyperfine state with $|m_F^\prime| = 1/2,~3/2,~5/2,~7/2$. 

Figure~\ref{figure_4}(d) displays the differential light shift for each value of $|m_F'|$, obtained from Lorentzian fits of the blow-out spectra, as a function of the tweezer intensity. These are compared with the measured differential light shifts in $^{174}$Yb. 
Through linear fits of the $^{174}$Yb data, we estimate a differential light shift of $0.029(3)\,\text{Hz cm$^2$/W}$ and $7.57(3)\,\text{Hz cm$^2$/W}$ between the ${}^1$S$_0$ ground state and the ${}^3$P$_1\,|J^\prime=1,m_J^\prime=0\rangle$ and $|J^\prime=1,m_J^\prime=\pm1\rangle$ states, respectively. These values are consistent with Ref.~\citenum{Saskin_2019}, and are then used to determine the expected differential light shifts for $^{173}$Yb sub-states, without relying on any \textit{ab initio} polarizability model (see Appendix~\ref{App:E}). In brief, we express the light shift of each of the $^3$P$_1\, |F'=7/2,m_F'\rangle$ states of $^{173}$Yb as a weighted sum of the measured light shifts of the $^3$P$_1\,|J^\prime=1,m_J^\prime=0,\pm1\rangle$ states of $^{174}$Yb, using the Clebsch-Gordan coefficients $|\langle F,m_F |\, J,m_J;I,m_I \rangle|^2$, with $F=7/2$, $J=1$, $I=5/2$ and $m_F=m_J+m_I$. 
We find that the measurements for $^{173}$Yb are in good agreement with the predicted light shifts, confirming the validity of our simple approach. Linear fits of the of $^{173}$Yb data for $|m_F'|=7/2$, $5/2$, $3/2$, and $1/2$ yield differential light shifts of $7.58(9)\,\text{Hz cm$^2$/W}$, $5.49(6)\,\text{Hz cm$^2$/W}$, $4.07(3)\,\text{Hz cm$^2$/W}$, and $3.27(3)\,\text{Hz cm$^2$/W}$, respectively.
Comparing the predictions with the fitted values, we find a discrepancy of less than $2\%$ for all states.

The complex internal structure of $^{173}$Yb, arising from its high nuclear spin $I = 5/2$, and the associated differential light shifts, introduce challenges for single-atom imaging. We optimize the cooling beams detuning to address atoms in both the $m_F = \pm1/2$ states. 
\fixme{To this end, a low B-field of about 0.2\,G is employed to avoid splitting significantly states with equal $|m_F'|$ in the $^3$P$_1$~$|F'=7/2\rangle$ level manifold.}
Yet, the cooling efficiency decreases for higher-$|m_F|$ states due to the quadratic light shift experienced by states with $|m_F'|> 1/2$. In particular, atoms in the $m_F = \pm 5/2$ ground states are essentially dark to the green cooling light, owing to the several-MHz differential light shift of the $^3$P$_1$~$|F'=7/2,m_F'=\pm3/2,~\pm5/2,~\pm7/2\rangle$ states. 
Employing a two-color imaging scheme allows to overcome this potential issue, and yields high detection fidelities with limited atom loss. 
The blue imaging light \fixme{is near-resonant with the ${}^1$S$_0\,|F=5/2\rangle \rightarrow {}^1$P$_1\,|F'=7/2\rangle$ transition} and acts as a continuous repumper, redistributing the ground state population from the dark $m_F = \pm 5/2$ states into the bright $m_F = \pm1/2,~\pm3/2$ states, thus improving the efficiency of green cooling. Moreover, the 5B MOT beams used as molasses feature a mix of polarizations that ensures $m_F$-symmetric optical pumping processes.
Specifically, \fixme{each of the horizontal beams} is set to an intensity of $I_{\text{HOR}} \simeq 3\,I_s$ and a detuning of $-8.5\,\text{MHz}$ from the light-shifted $^3$P$_1 \ket{F'=7/2, m_F'=\pm1/2}$ states, while the vertical beam has an intensity of $I_{\text{VER}} \simeq 1.6\,I_s$ and a detuning of $-6.5\,\text{MHz}$ from the same states. 
\fixme{We find that the loss probability is weakly sensitive to the exact cooling beam detuning in a few-MHz range around the optimal point.}

In Fig.~\ref{figure_5}(a) we present the extracted array-averaged single-atom imaging infidelity and survival probability after a $50\,\text{ms}$ imaging pulse, as a function of the imaging beam intensity. 
At the intensity indicated by the vertical shaded band, $I_{399} \simeq 1.3\times10^{-2}\,I_s$, we collect $36.8(1)$ photons per atom on average,
resulting in a fidelity of $99.96(1)\%$ and a survival probability of $98.5(2)\%$. 
\fixme{At high intensities, the survival probability decreases while the infidelity saturates around $4\times10^{-4}$, approaching the lower limit set by the number of recorded images.}
The inset presents a histogram of the photon counts in 3\,px $\times$ 3\,px regions of interests (ROIs), where two peaks can be clearly distinguished, one corresponding to the background signal and the other to the single-atom signal. 
To characterize the imaging performances we acquire two successive images separated by $50\,\text{ms}$. Figure~\ref{figure_5}(b) shows the correlation between the photon counts in the first and second image for $I_{399} \simeq 1.3\times 10^{-2}\,I_s$.  
\fixme{For each value of the imaging intensity, the single-atom detection fidelity and survival probability are determined following the approach outlined in Appendix~\ref{App:F}}.
The gray dashed lines in Fig.~\ref{figure_5}(b) represent the optimized photon-count thresholds that maximize the global fidelity. To obtain error bars for the fidelity and survival probability we employ a bootstrapping procedure~\cite{Holland_2023}.

\section{Conclusions}
In this work, we have demonstrated high-fidelity and low-loss single-atom imaging of an ytterbium tweezer array loaded from a 5B MOT. We have shown that this MOT configuration allows to prepare cold and dense samples with short loading times for the most common ytterbium isotopes, making it a viable and practical solution for quantum science platforms employing high-NA objective lenses. \fixme{The 5B MOT approach may be adapted to other AEAs atoms, even for the extremely narrow natural ${}^1\text{S}_0-{}^3\text{P}_1$ linewidths found in Ca, Mg, or Be, by employing multifrequency MOTs~\cite{Kuwamoto_1999, Snigirev_2019} and/or by incorporating an appropriate repumping scheme to gain control on the ${}^3$P$_1$ state lifetime and branching off to dark states~\cite{Leibfried_2003,Zhang_2022}.}
\fixme{Moreover, while our implementation uses a single microscope objective, this approach is compatible with confocal, two-objective assemblies. Although integrating a second objective introduces additional complexity, the 5B MOT configuration remains applicable, for instance by directing the vertical MOT beam from below and collimating it with the bottom objective.}

\fixme{Addressing complications from its large-spin structure and} non-magic trapping wavelength, we have demonstrated the first high-fidelity single-atom imaging of $^{173}\text{Yb}$ atoms.
Our two-color imaging scheme can be adapted to other fermionic AEAs, which lend themselves to high-dimensional quantum encodings exploiting their large and robust nuclear spin \cite{Omanakuttan_2021,Cuadra_2022,Ahmed_2025}. Moreover, the demonstrated single-atom detection scheme may be directly extended to quantum gas microscopes where, complemented by the nuclear spin-sensitive control of the metastable $^3$P$_0$ clock state, it will enable the investigation of two-orbital and/or SU($N$)-symmetric lattice models with single-particle resolution \cite{Gorshkov_2010,Cazalilla_2014,KanaszNagy_2018,Riegger_2018,Hofrichter_2016,Pasqualetti_2024}.

\begin{acknowledgments}
We thank M.~Aidelsburger, N.~Bruno, G.~Cappellini, N.~Darkwah Oppong, F.~Ferlaino, T.~H\"ohn, G.~Roati and L.~Tarruell for useful discussions, and M.~Marinelli for a careful reading of this manuscript. We also thank G.~Brajnik, G.~Cautero, A.~Fondacaro, A.~Martin, F.~Salvador, A.~Trenkwalder and the Instrumentation and Detectors Laboratory of Elettra Sincrotrone for technical support during the construction of the experimental apparatus, and F.~Barbuio, G.~Bartolini, G.~Tazzoli, S.~Vigneri,  and J.~Wang for participating in the initial setting up of the experiment. 
This work has received financial support from the European Research Council (ERC) under the European Union’s Horizon 2020 research and innovation programme (project OrbiDynaMIQs, GA No.~949438), and from the Italian MUR under the FARE 2020 programme (project FastOrbit, Prot.~R20WNHFNKF). This work has also received funding from the European Union under the Horizon Europe program HORIZON-CL4-2022-QUANTUM-02-SGA (project PASQuanS2.1, GA no.~101113690), and by the Next Generation EU (Missione 4, Componente 1) under the MUR PRIN 2022 programme (project CoQuS, Prot.~2022ATM8FY) and the PNRR MUR project PE0000023-NQSTI. 

\end{acknowledgments}

\smallskip
\paragraph*{Author contributions---} O.A.K., A.M.F., R.P., W.L. and F.S. conceptualized the work and conducted the experiments. O.A.K. and A.M.F. analyzed the data and performed numerical simulations under the supervision of F.S.. O.A.K., A.M.F. and F.S. wrote the manuscript. 

\smallskip
The authors declare no competing interests.

\appendix
\renewcommand{\appendixname}{APPENDIX}
\section{EXPERIMENTAL SEQUENCE}\label{App:A}

\begin{figure}[t!]
	\begin{center}
		\includegraphics[scale=0.8]{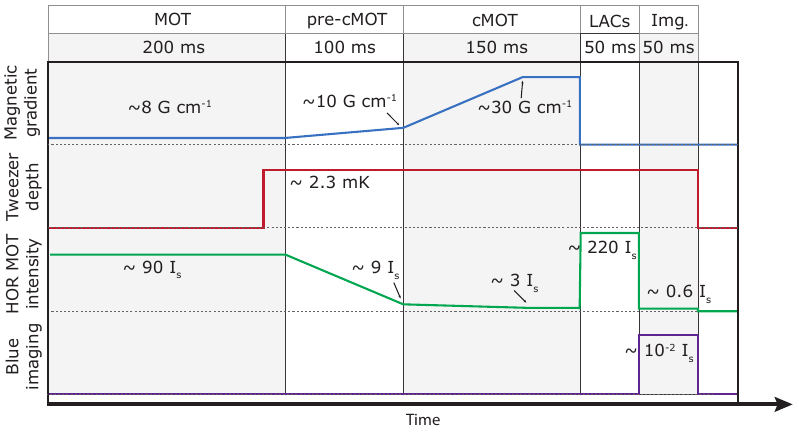}
		\caption{Time diagram of the experimental sequence for trapping and imaging single $^{173}\text{Yb}$ atoms in optical tweezers. The sequence begins with a $200\,\text{ms}$ MOT loading phase, followed by a $250\,\text{ms}$ compression ramp, divided in two cMOT stages. The tweezers are turned on a few milliseconds before the first compression stage. After the cMOT is turned off, we apply a $50\,\text{ms}$ LACs pulse to obtain single atoms in the tweezers. The two-color imaging typically lasts $50\,\text{ms}$.}
    \label{figure_6}
	\end{center}
\end{figure}

Our experimental apparatus is based on a commercial atomic source (AOSense Yb Beam RevC), featuring an integrated permanent magnet Zeeman slower (ZS) and a 2D MOT stage. A collimated hot ytterbium beam is generated by an oven kept at a temperature of $380^\circ\text{C}$, producing a flux of approximately $1.5 \times 10^{12}\,\text{atoms/s}$. The ZS is employed to reduce the longitudinal atomic beam velocity to approximately $40\,\text{m/s}$. The 2D MOT is utilized to cool atoms in the transversal direction and to deflect the beam towards an octagonal glass cell, where all experiments are performed. The typical experimental sequence employed for trapping and imaging single atoms in optical tweezers is illustrated in Fig.~\ref{figure_6}. The sequence begins by activating the 2D MOT and the slowing crossed beams (CBs, see main text) together with the green MOT beams and magnetic-field gradient to load the narrow-line MOT. After typically $200\,\text{ms}$ of loading, the 2D MOT beams are turned off to prevent any residual flux from reaching the glass cell. Simultaneously the CBs are turned off while the ZS remains continuously on during the experiment. The MOT loading is followed by a two-stage compression lasting approximately $250\,\text{ms}$. \fixme{Compression in the 5B MOT follows a similar procedure to that used in a 6B configuration: the laser power and detuning are gradually reduced while the magnetic field gradient is increased. 
The main difference between the 6B and 5B configurations lies in the vertical beam parameters adjustment: since its power is already low during the MOT loading, only a minimal reduction is required during compression. 
The first stage (pre-cMOT) involves a slow compression of the atom cloud, during which the field gradient is increased to $10\,\text{G/cm}$. 
At the same time, the power and global detuning of both the horizontal and vertical MOT beams are reduced. 
In the subsequent final cMOT stage, the magnetic field gradient is ramped up to approximately $30\,\text{G/cm}$ over a linear ramp of $100\,\text{ms}$, while the MOT beams' powers and detuning are further adjusted to yield the lowest temperatures.}
The following $50\,\text{ms}$ are used to move the cloud over the tweezers position. The tweezer traps are turned on a few milliseconds before the pre-cMOT phase, with a typical trap depth of $2\,\text{mK}$. After the cMOT phase, the quadrupole magnetic field is switched off, and a $50\,\text{ms}$ light-assisted collisions (LACs) pulse is applied to isolate single atoms in the tweezers. Imaging follows the LACs phase, with a typical duration of $50\,\text{ms}$. 

\section{FIVE-BEAM MOT SIMULATIONS} \label{App:B}
\begin{figure}[b!]
    \centering
    \includegraphics[scale=0.5]{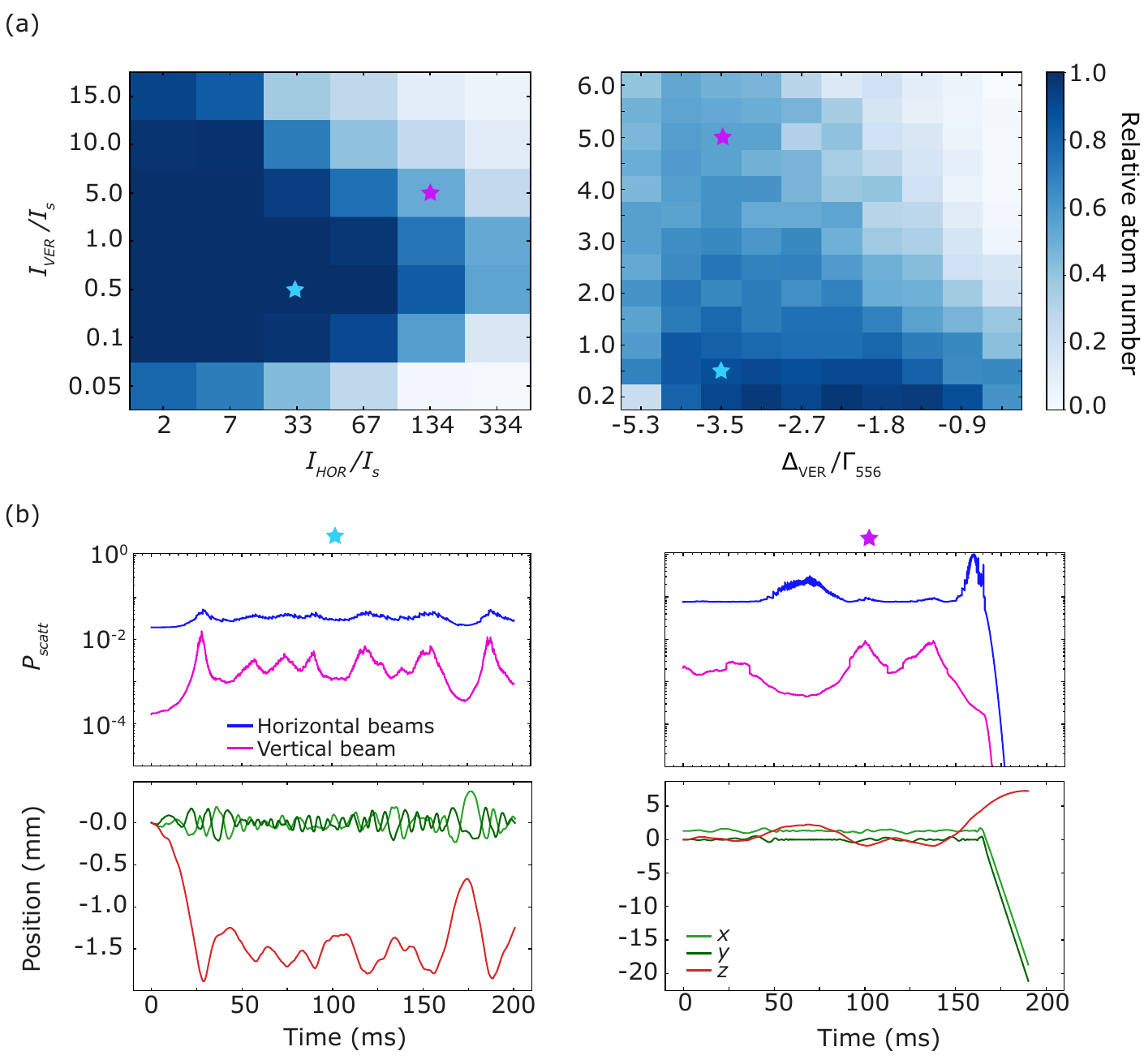}
    \caption{Monte Carlo simulations of the five-beam MOT. (a) Fraction of atoms trapped in the MOT as a function of the vertical and horizontal beam intensities and the vertical beam detuning. Atoms are initialized at the intersection of the MOT beams, corresponding to the center of the quadrupole field, with zero velocity. An atom is considered trapped if it is retained in the MOT region after $200\,\text{ms}$ of time evolution. Each point is obtained from $100$ individual trajectories. The cyan and purple stars indicate representative points of the 5B MOT operation, corresponding to high and low MOT stability respectively. 
    (b) Illustrative scattering probability (top) and associated atomic trajectory (bottom) as a function of time for the two scenarios marked by stars in panel (a). Blue (magenta) curves represent the probability of scattering from any of the horizontal beams (the vertical beam) during the simulation. 
    For the first condition (cyan star), where the trapped fraction is maximized, the atom experiences repeated bounces in the vertical direction caused by sudden variations in the vertical beam scattering probability. For the second condition (purple star), where intensities are higher and the trapped fraction is lower, the atom is pushed upward by the vertical beam and eventually escapes horizontally after approximately $100\,\text{ms}$.
    }
    \label{figure_51}
\end{figure}
Monte Carlo simulations offer valuable insight into the working principle of the 5B MOT. Here, we present additional simulations beyond those in the main text to provide a more comprehensive overview of the 5B MOT operation. Figure~\ref{figure_51}(a) illustrates the trapped fraction in the MOT after $200\,\text{ms}$ of time evolution as result of multiple simulated single-atom trajectories. Atoms are initialized at the intersection of the MOT beams, corresponding to the center of the quadrupole field, and evolve under the combined influence of the MOT beams, magnetic field gradient and gravity. We assume a textbook spin structure of the $^1$S$_0$ and $^3$P$_1$ states, such as in $^{174}$Yb ($I=0$). The simulation accounts only for losses arising from the atomic motion out of the trap, and it does not include e.g.~background gas-induced or density-induced losses.
Each trajectory is divided into steps of duration $\tau = 1/\Gamma$, during which the atom can scatter a photon from one of the MOT beams.
In particular, at each step we evaluate the scattering probability for each MOT beam as $P_{i} = \tau \cdot \Gamma_{sc,i}$, where $\Gamma_{sc,i}$ is the scattering rate of beam $i$ and depends on the beam parameters as well as on the atomic position and velocity. For each beam independently, we stochastically determine if the scattering event is successful. If more than one beam yield a scattering event, we randomly choose one of them with uniform probability. After determining whether the atom has scattered a photon, and from which beam, we update its velocity taking into account absorption and isotropic spontaneous emission and evolve its coordinates with an equation of motion before proceeding to the next time step.
For each data point in the figure, we simulate $100$ trajectories and we determine the trapped fraction as the fraction of atoms in the MOT region at the end of the time evolution. We first investigate the interplay between the vertical and horizontal beams intensity, finding that the vertical beam requires significantly lower intensity than \fixme{each} horizontal beam. Moreover, we observe a correlation between the vertical and horizontal beams power: as the vertical beam power is changed from the optimal value, the MOT becomes less robust and requires less power in the horizontal beams.
By fixing the horizontal beams intensity we also observe a correlation between the vertical beam parameters. In particular, we find that larger detunings require larger intensities to hold the atoms against gravity, as expected from the inverse dependence on detuning of the scattering force.  
The optimal working point obtained in simulations is in agreement with experimental results for $^{171}$Yb shown in Fig.~\ref{figure_2}(a) except for a small difference in the intensity of the vertical beam. We attribute this difference to the internal structure of the fermionic isotope, which is not accounted for in our simulations. In Fig.~\ref{figure_51}(b), we present two illustrative trajectories to provide insight into the results shown in Fig.~\ref{figure_51}(a). In particular we show the scattering probability from the horizontal and vertical beams, along with the atom position as a function of time, for two representative simulation scenarios: one where the captured fraction is maximized and another one where it is around $50\%$. 
For the first one, the scattering probability from the vertical beam is generally much lower than that of the horizontal beams, which can be attributed to lower power broadening of the transition compared to the more intense horizontal beams. 
Indeed, the atoms do not scatter many photons from the vertical beam until the resonant condition is matched. This is marked by a peak in the scattering probability and a sudden inversion of motion along the vertical direction. This results in the peculiar shape of the atomic vertical trajectory, characterized by a series of free falls and bounces which keep the atoms inside the MOT region. 
When the vertical intensity is too high, atoms have a higher probability of escaping. As visible from the second trajectory, this occurs because the atom is pushed upward by the vertical beam until it enters a region of the magnetic quadrupole field where where horizontal confinement is less robust. Lowering the horizontal beam power in this case is beneficial, as it reduces the horizontal scattering rate, giving the atom more time to fall back in a more stable region of the MOT before scattering from the horizontal beams.

\section{TWEEZER ARRAY HOMOGENIZATION} \label{App:C}
Imaging performances are sensitive to amplitude inhomogeneities across the tweezer array. Inhomogeneities introduce variations in the ${}^1$S$_0 - {}^3$P$_1$ differential light shift, leading to a reduced cooling efficiency in our imaging protocol. To achieve uniform trapping across the array, we apply two amplitude homogenization steps, similarly to the method described in Ref.~\citenum{Jenkins_2022}. In the first step, we acquire an image of the optical tweezer array using a CMOS camera (Zelux CS165MU) and fit the image with a sum of Gaussian functions. For each tweezer we determine an error, defined as: 
\begin{equation}
E_i = \frac{I_i - \langle I\rangle}{\langle I\rangle}
\end{equation}
where  $I_i$ is proportional to the intensity of each tweezer and $\langle I \rangle$ is the average of the $I_i$ values. 
The amplitudes $A_i$ of the new RF tones driving the AOD are obtained by taking the difference between the old RF array and the retrieved error, adding a proportional gain:

\begin{equation}
A_i^{new} = A_i^{old} - p E_i
\end{equation}

\begin{figure}[b!]
    \begin{center}
    \includegraphics[scale=0.5]{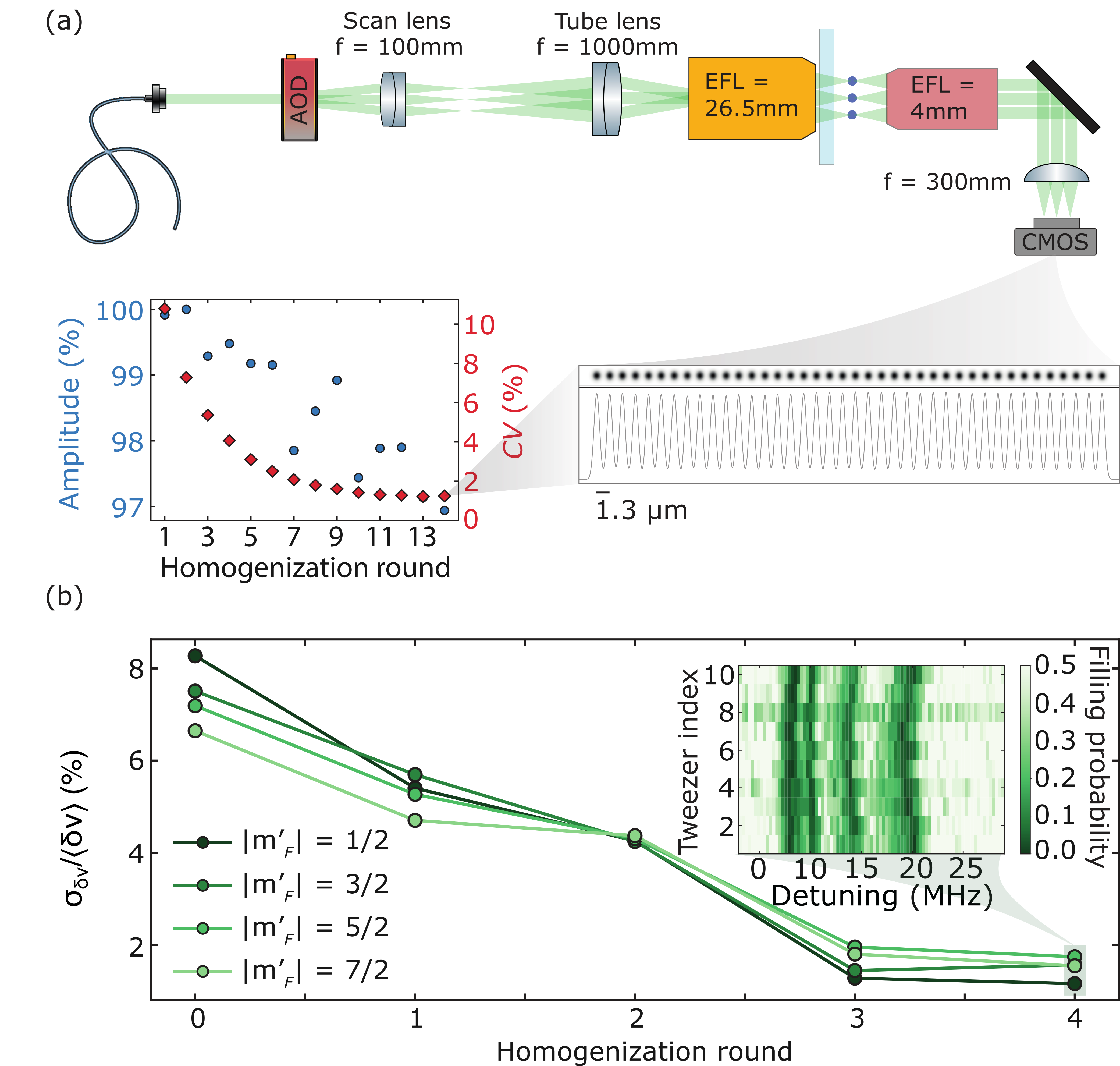}
		\caption{Tweezer amplitude homogenization. (a) Optical tweezers are produced using an AOD. The intermediate tweezers, generated using a scan lens with $f=100\,\text{mm}$, are imaged onto the atomic plane through a tube lens with $f=1000\,\text{mm}$ and the microscope objective. The intensity distribution of each tweezer trap at the atom plane is measured using another objective with $\text{EFL}=4\,\text{mm}$ and a tube lens with $f=300\,\text{mm}$, focusing on a CMOS camera. The plot shows the mean amplitude and coefficient of variation (CV) across a $40$-tweezer array, spaced by $1.3$\,$\mu$m, as a function of feedback iterations. The CV reaches $1.4\%$ after $14$ iterations. The inset shows the image and the fit of the tweezer array after the final homogenization step. (b) Differential light shifts throughout the tweezer array are measured via blow-out spectroscopy and used to correct intensity inhomogeneities. The plot shows the \fixme{relative RMS variation of measured light-shifts} as a function of iteration round for a $10$-tweezer array with $\sim 8\,\mu$m spacing. The inset shows the blow-out spectroscopy results after the final homogenization round.}
        \label{figure_8}
	\end{center}
\end{figure}

The feedback loop is iterated until the amplitude homogeneity converges to a stationary and satisfactory level.
We can perform this feedback protocol either imaging the tweezers produced by the microscope objective at the atomic plane, or the intermediate tweezers produced by the scan lens (see scheme in Fig.~\ref{figure_8} (a)). 
Imaging the tweezers at the atomic plane, yields a remarkably low coefficient of variation (CV) of $1.4\%$ across an array of $40$ tweezer spaced by $1.3\,\mu\text{m}$ after $14$ iterations, as shown in Fig.~\ref{figure_8}(a). 
However, directly imaging the tweezers at the atom plane is not always practical.
Therefore, we implement the feedback loop at the scan lens focal plane, upstream of the tube lens and the objective. At this stage, the feedback loop corrects for inhomogeneities arising from nonlinearities in the AOD crystal and RF amplifier. While this approach improves the uniformity at the scan lens plane, it does not account for aberrations introduced by downstream optics, and thus does not guarantee optimal uniformity at the atom plane. Nonetheless, following the homogenization, we measure a CV of $5\%$ at the atom plane, making a significant improvement from the initial CV of $11\%$. To further reduce the CV, we apply a second homogenization procedure that leverages light shifts in non-magic traps to finely compensate any remaining inhomogeneities.

To this end, we exploit the differential light shifts induced on the ${}^1$S$_0 \rightarrow {}^3$P$_1$ transition by the $532\,\text{nm}$ trapping light for $^{173}$Yb. We apply a similar amplitude homogenization protocol as before, but use the atomic light shift as the figure of merit in this case. 
To measure the light shifts we perform blow-out spectroscopy on the ${}^1$S$_0 \rightarrow {}^3$P$_1$ transition using a green pulse to expel atoms from the trap, followed by a blue imaging pulse. The blow-out pulse is linearly polarized at $45^\circ$ with respect to the quantization axis, allowing excitation of both the $\pi$ and $\sigma^{\pm}$ transitions. This allows us to resolve four distinct transitions connecting the ${}^1$S$_0 \ket{m_F}$ states to the ${}^3$P$_1 \ket{m_F' = \pm1/2,~\pm3/2,~\pm5/2,~\pm7/2}$ states. By measuring the differential light shift in each tweezer, we can adjust the AOD driving RF tone amplitudes to correct for inhomogeneities.
\fixme{In Fig.~\ref{figure_8}(b), we show the homogenization results, plotting the ratio between the standard deviation and the average of the measured differential light shifts across the tweezer array.} We achieve a \fixme{relative variation} of $1.1\%$ for the ${}^1$S$_0 \rightarrow {}^3$P$_1 \ket{F'=7/2, m_F'=\pm1/2}$ transitions, which are the ones we address for molasses cooling during imaging. The inset of Fig.~\ref{figure_8}(b) shows the results of the blow-out spectroscopy performed after the final round of homogenization. \fixme{It is important to note that the residual inhomogeneity of the atomic spectroscopic signal reflects variations in both trap depth and tweezer waist, making it not directly comparable to the CV obtained from optical amplitude measurements.}

\section{TWEEZER WAIST CHARACTERIZATION}\label{App:D}
A crucial parameter to characterize an optical tweezer trap is the trap frequency. We determine the tweezer trap frequencies using the parametric heating 
technique \cite{Joykutty_2005}. 
This method relies on periodically modulating the trap depth to induce atom losses, which are maximized when the modulation frequency is twice the trapping frequency. We begin by loading a ten-site optical tweezer array at a trap depth of $2\,\text{mK}$ and take a low-loss initial image.
The trap depth is then lowered to about $1\,\text{mK}$ to enhance atom loss, followed by amplitude modulation for $100\,\text{ms}$. After modulation, the trap depth is ramped back to $2\,\text{mK}$, and a second image is taken to count the remaining atoms. The frequency spectrum reveals two distinct resonances, corresponding to the axial and radial trap frequencies. The presence of a single radial resonance indicates the absence of astigmatism in the optical tweezer traps. 
From the measured trap frequencies, we indirectly determine the tweezer waist by expressing it in terms of the ratio of the radial to axial trap frequencies. This approach removes dependence on the trap depth, allowing the tweezer waist to be written as:
\begin{equation} 
w_0 = \frac{\lambda}{\pi \sqrt{2}}\frac{\omega_r}{\omega_z}
\end{equation}
Using this relation along with the experimentally measured radial and axial trap frequencies, we determine the tweezer waist to be: $w_0 = 578(4)\,\text{nm}$. 
This value is obtained after correcting misalignments of the objective lens relative to the glass cell, which introduce aberrations and lead to a reduction in both the trap depth and trap frequencies. To this end, we adjust the objective tilt in order to maximize the axial trap frequency measured with parametric heating. 

\section{POLARIZABILITY CALCULATIONS}\label{App:E}
The optical potential experienced by an atom is determined by the product of the state-dependent polarizability and the intensity profile of the trap:
\begin{equation}
\label{acStark}
U(\textbf{r}) = -\frac{1}{2\epsilon_{0}c}\alpha(\omega)I(\textbf{r})
\end{equation}
The polarizability $\alpha$ of state $\ket{i}$ is obtained from second-order perturbation theory \cite{Grimm_2000}:
\begin{equation}
\alpha(\omega) = \sum_{f}\frac{2\omega_{if}|\bra{i}\pmb{\epsilon}\cdot \textbf{d}\ket{f}|^{2}}{\hbar(\omega_{if}^{2}-\omega^{2})}
\end{equation}
where \textbf{d} is the dipole operator that connects the atomic level of interest $\ket{i}$ to upper atomic levels $\ket{f}$, $\pmb{\epsilon}$ is the polarization of the light and $\hbar\omega_{if}$ is the energy difference between state $\ket{i}$ and $\ket{f}$. \fixme{To treat angular-momentum structure of atomic states, the polarizability must be generalized to a polarizability tensor}. For the atomic state $\ket{i}$, this can be expressed as a Cartesian tensor as \cite{Steck_2011}:
\begin{equation}
\alpha_{\mu\nu}(\omega) = \sum_{f}\frac{2 \omega_{if}}{\hbar (\omega_{if}^2 - \omega^2)} \bra{i}\hat{d}_\mu \ket{f}\bra{f}\hat{d}_\nu\ket{i}.
\end{equation}
To relate the transition dipole moments to experimentally measured lifetimes, we use the Wigner-Eckart theorem, which allows to express the dipole moment as the product between a scalar quantity, i.e.~the so-called reduced matrix element, and the Clebsch-Gordan coefficients \cite{Steck_2011}:
\begin{equation}
\label{dipole_moment}
\begin{split}                     
&\bra{n, J, m_J}\hat{d}_q\ket{n', J', m_J'} =\\ 
&=\bra{n, J}\hat{d}\ket{n', J'} (-1)^{J'-1-m_J}  \sqrt{2J+1}\begin{pmatrix}
1 & J' & J \\
q & m_J' & -m_J
\end{pmatrix}\\[2mm]
\end{split}
\end{equation}
where the quantum number $n$ labels the electronic configuration, $J$ is the total electronic angular momentum and $m_{J}$ is the angular momentum projection along the quantization axis. The polarization of the light is defined by the index $q$, with $q = \pm 1$ corresponding to right- and left-circular polarization, respectively, and $q = 0$ to linear polarization along the quantization axis. The term in parenthesis is the Wigner 3-j symbol. The reduced matrix element $\bra{n, J}\hat{d}\ket{n', J'}$ can be related to the decay rate of the $nJ'$ states:
\begin{equation}
\nonumber                    
\Gamma_{n'J',nJ} = \frac{\omega_{n'J'nJ}^3}{3\pi\hbar c^3\epsilon_{0}}\frac{|\bra{n, J}\hat{d}\ket{n', J'}|^2}{2J'+1}
\end{equation}
where $\omega_{n'J'nJ}$ is the transition angular frequency. This approach allows us to determine the polarizability of $\ket{J,m_J}$ states from the measured lifetimes of the relevant excited states with electronic quantum number $n'$ and total electronic angular momentum $J'$. We apply this method to calculate the polarizability of the $^1\text{S}_0 \ket{J=0,m_J=0}$ and $^3\text{P}_1 \ket{J=1, m_J = 0,\pm 1}$ states of ${}^{174}\text{Yb}$. This model does not account for any hyperfine structure, and is thus valid for $I=0$. 
On the other hand, hyperfine structure becomes important for fermionic isotopes, where $I\neq0$ and the total atomic angular momentum, $F \neq J$ and its projection $m_F$ are good quantum numbers. 

To extend our model to fermionic isotopes, and in particular to ${}^{173}\text{Yb}$, we follow a simple approach. Since the $\ket{F,m_F}$ states are given by superpositions of $\ket{J,m_J;I,m_I}$ states, whose amplitudes are the Clebsch-Gordan coefficients $\langle F,m_F |\, J,m_J;I,m_I \rangle$, we can express the polarizability $\alpha_{\ket{m_F}}$ of the $^3\text{P}_1\ket{F=7/2, m_F}$ states of ${}^{173}\text{Yb}$ as:
\begin{equation}
\begin{split}
\alpha_{\ket{m_F=\pm7/2}} &= \alpha_{\ket{m_J=\pm1}} \\
\alpha_{\ket{m_F=\pm5/2}} &= \frac{5}{7} \alpha_{\ket{m_J=\pm1}} +  \frac{2}{7} \alpha_{\ket{m_J=0}} \\
\alpha_{\ket{m_F=\pm3/2}} &= \frac{11}{21} \alpha_{\ket{m_J=\pm1}} +  \frac{10}{21} \alpha_{\ket{m_J=0}} \\
\alpha_{\ket{m_F=\pm1/2}} &= \frac{3}{7} \alpha_{\ket{m_J=\pm1}} +  \frac{4}{7} \alpha_{\ket{m_J=0}}
\end{split}
\label{polar_combination}
\end{equation}
where $\alpha_{\ket{m_J}}$ is the polarizability of the $^3\text{P}_1~\ket{J=1, m_J}$ state of ${}^{174}\text{Yb}$.
We thus extend our polarizability model to fermionic isotopes without explicitly including the total atomic angular momentum $F$ and its projection $m_F$ in the calculation of the dipole matrix elements. Thereby, we can predict the light shifts of the $^3\text{P}_1\ket{F=7/2,m_F}$ states in ${}^{173}\text{Yb}$ based on the measured light shifts of the ${}^{174}\text{Yb}$ $^3\text{P}_1\ket{J=1, m_J}$ states [see Fig.~\ref{figure_4}(d)]. 

In our model, the polarizabilities of the $^1\text{S}_0$, $^3\text{P}_1$, and $^3\text{P}_0$ states are corrected to match known constraints. In particular, the asymptotic value of the $^1\text{S}_0$ state polarizability is reduced by $-1.4$\,Hz\,cm$^2$/W to align with the static polarizability value calculated in \cite{Dzuba_2010}. To match the experimentally determined magic wavelength of $532\,\text{nm}$ for the $^1\text{S}_0$ $|J=0, m_J=0\rangle \rightarrow {}^3\text{P}_1|J'=1, m_J'=0\rangle$ transition in $^{174}$Yb, a small offset of $-0.15$\,Hz\,cm$^2$/W is applied to the polarizability of the $^3\text{P}_1$ state. This adjustment also correctly predicts additional magic wavelengths, including $487\,\text{nm}$ for the $^3\text{P}_1|F=3/2, |m_F|=1/2\rangle$ state in $^{171}$Yb, which is experimentally measured in \cite{Ma_2022}. With these corrections, we find that the differential light shift between the $^1\text{S}_0$ $|J=0, m_J=0\rangle$ and ${}^3\text{P}_1 |J'=1, m_J'=\pm1\rangle$ states of $^{174}$Yb lies within $20\%$ from our experimental measurements. This remaining discrepancy likely arises from inaccuracies in the tweezer waist estimation and power calibration.

\fixme{\section{CHARACTERIZATION OF SINGLE-ATOM IMAGING PERFORMANCE}\label{App:F}

\begin{figure*}[t!]
    \centering
    \includegraphics[width=0.65\textwidth]{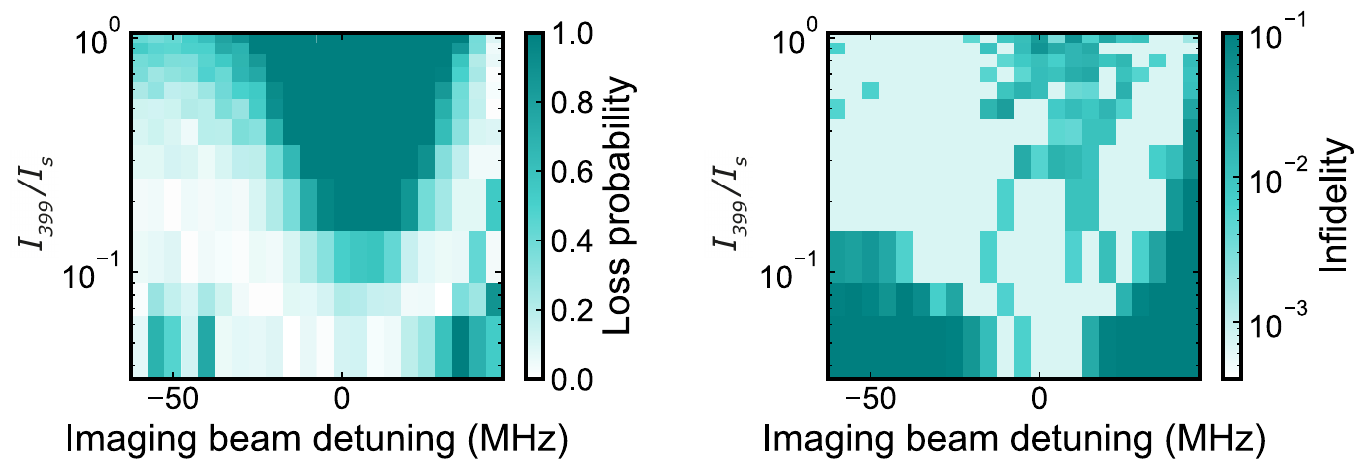}
    \caption{\fixme{Dependence of single-atom ${}^{173}\text{Yb}$ imaging performance on the blue imaging beam parameters. Atom-loss probability (left) and single-atom detection infidelity (right) as a function of imaging beam intensity and detuning from the ${}^1\text{S}{}_0 \rightarrow {}^1$\text{P}${}_1|F'=7/2\rangle$ transition. The imaging pulse duration is set to $50\,\text{ms}$ and each point is obtained from $\sim280$ individual shots.}}
    \label{figure_9}
\end{figure*}
To estimate the single-atom detection fidelity of our ${}^{173}\text{Yb}$ single-atom imaging scheme and the associated losses we employ a model-free approach based on acquiring two equal images in succession~\cite{Norcia_2018, Holman_2024}. 
In particular, we analyze the photon counts of the two images, set a threshold for distinguishing atom occupancy and determine the probabilities of identifying an atom in both images, a void in both images, an atom and then a void or a void and then an atom. These probabilities depend on the detection fidelity and the atom survival probability between images, as well as on the initial loading fraction. 
For each set of imaging parameters, we compute these quantities employing different values of the photon-count threshold that discriminates between atoms and voids, and we choose the value that maximizes the fidelity.

Following this procedure, we find a set of imaging parameters, marked by the shaded line in Fig.~\ref{figure_5}(a), that provides the best trade-off between low losses and high fidelity.
For our parameters of choice, we observe a $1.5\%$ atom-loss probability per image, which arises from multiple contributions. 
In particular, the most significant loss mechanisms are off-resonant scattering of trap photons from the ${}^1\text{P}_1$ excited state, which can lead to excitation to anti-trapped states as well as to two-photon photoionization events, and the finite lifetime of atoms undergoing cooling in tweezers.
These two contributions add up to approximately $0.9\%$ loss probability. 
Additionally, decay from ${}^1\text{P}_1$ to triplet dark states and losses due to vacuum background collisions lead to $0.04\%$ loss probability.
The remaining $0.55\%$ is ascribed to cooling inefficiencies, likely caused by non-magic trapping and the complex structure of ${}^{173}\text{Yb}$.

In Fig.~\ref{figure_9} we show the dependence of the imaging performance on the intensity and detuning of the blue imaging beam for fixed illumination time of $50\,\text{ms}$. 
Both the imaging loss probability and the detection infidelity show a characteristic V-shaped dependence on intensity and detuning from resonance, indicating that the dominant factor is the number of scattered blue photons. 
In general, while a large detuning allows to work with larger intensity, the red-detuned side displays a greater robustness, likely due to reduced heating compared to the blue-detuned side as well as to possible sub-Doppler cooling effects from the retroreflected imaging beam itself~\cite{Kostylev_14}. For the results presented in Fig.~\ref{figure_5}, we operate with a small detuning of -5\,MHz from resonance.}

\bibliographystyle{apsrev4-1_ourstyle}
\bibliography{Yb173_5B.bib}

\end{document}